\PassOptionsToPackage{unicode}{hyperref}
\PassOptionsToPackage{hyphens}{url}
\documentclass[
  11pt,
]{article}
\usepackage{xcolor}
\usepackage[margin=1in]{geometry}
\usepackage{amsmath,amssymb}
\usepackage{iftex}
\ifPDFTeX
  \usepackage[T1]{fontenc}
  \usepackage[utf8]{inputenc}
  \usepackage{textcomp} % provide euro and other symbols
\else % if luatex or xetex
  \usepackage{unicode-math} % this also loads fontspec
  \defaultfontfeatures{Scale=MatchLowercase}
  \defaultfontfeatures[\rmfamily]{Ligatures=TeX,Scale=1}
\fi
\usepackage{lmodern}
\ifPDFTeX\else
\fi
\IfFileExists{upquote.sty}{\usepackage{upquote}}{}
\IfFileExists{microtype.sty}{% use microtype if available
  \usepackage[]{microtype}
  \UseMicrotypeSet[protrusion]{basicmath} % disable protrusion for tt fonts
}{}
\makeatletter
\@ifundefined{KOMAClassName}{% if non-KOMA class
  \IfFileExists{parskip.sty}{%
    \usepackage{parskip}
  }{% else
    \setlength{\parindent}{0pt}
    \setlength{\parskip}{6pt plus 2pt minus 1pt}}
}{% if KOMA class
  \KOMAoptions{parskip=half}}
\makeatother
\NewDocumentCommand\citeproctext{}{}
\NewDocumentCommand\citeproc{mm}{%
  \begingroup\def\citeproctext{#2}\cite{#1}\endgroup}
\makeatletter
 \let\@cite@ofmt\@firstofone
 \def\@biblabel#1{}
 \def\@cite#1#2{{#1\if@tempswa , #2\fi}}
\makeatother
\newlength{\cslhangindent}
\newlength{\csllabelwidth}
\newenvironment{CSLReferences}[2] % #1 hanging-indent, #2 entry-spacing
 {\begin{list}{}{%
  \setlength{\itemindent}{0pt}
  \setlength{\leftmargin}{0pt}
  \setlength{\parsep}{0pt}
  \ifodd #1
   \setlength{\leftmargin}{\cslhangindent}
   \setlength{\itemindent}{-1\cslhangindent}
  \fi
  \setlength{\itemsep}{#2\baselineskip}}}
 {\end{list}}
\usepackage{calc}

\providecommand{\tightlist}{%
  \setlength{\itemsep}{0pt}\setlength{\parskip}{0pt}}
\usepackage{amsmath}
\usepackage{amssymb}
\usepackage{hyperref}
\usepackage{bookmark}
\IfFileExists{xurl.sty}{\usepackage{xurl}}{} % add URL line breaks if available
\hypersetup{
  pdftitle={ssys: Exact algebraic recasting of ODE models into S-system or GMA form},
  pdfauthor={William S. Hlavacek},
  hidelinks,
  pdfcreator={LaTeX via pandoc}}

\title{ssys: Exact algebraic recasting of ODE models into S-system or
GMA form}
\author{William S. Hlavacek}
\date{2026-07-06}

\begin{document}
\maketitle

\section{Summary}\label{summary}

\texttt{ssys} is a Python toolkit for exact algebraic transformation of
ordinary differential equation (ODE) models into S-system or Generalized
Mass Action (GMA) form. Given a model in Antimony format
(\citeproc{ref-Heydarabadipour2026Antimony}{Heydarabadipour et al.,
2026}; \citeproc{ref-Smith2009Antimony}{Smith et al., 2009}), or an SBML
file (\citeproc{ref-Keating2020SBML}{Keating et al., 2020}) parsed with
\texttt{ssys.parse\_sbml}, \texttt{ssys} produces a mathematically
equivalent representation where each equation has one of the following
forms:

\textbf{S-system form} (difference of two monomials): \begin{equation}
\dot X_i = \alpha_i \prod_j X_j^{g_{ij}} - \beta_i \prod_j X_j^{h_{ij}}, \quad \alpha_i,\beta_i \geq 0 .
\label{eq:ssystem}
\end{equation}

Equation \ref{eq:ssystem} is a relaxed S-system form that permits
zero-valued right-hand-side terms; canonical S-system form requires both
constant coefficients in every equation to be strictly positive.

\textbf{GMA form} (sum of monomials): \begin{equation}
\frac{dX_i}{dt} = \sum_k \gamma_{ik} \prod_j X_j^{f_{ijk}} .
\label{eq:gma}
\end{equation}

The transformation introduces auxiliary variables as needed to decompose
a broad class of nonlinearities---including rational functions,
exponentials, logarithms, sine, and cosine---into products of power-law
terms. The recast is exact: the original and transformed systems have
identical dynamics on the invariant constraint manifold defined by
auxiliary variable definitions, given consistent initial conditions.
Exactness is local to trajectories in the admissible real domain:
denominators must avoid zero, logarithm arguments must remain positive,
square-root arguments must remain valid, lifted variables used as
power-law bases must remain positive, and auxiliary initial conditions
must satisfy their defining relations.

The software provides a command-line interface (CLI) for batch
processing Antimony models, a three-test validation suite (symbolic,
numerical, trajectory), and generates Jupyter notebooks
(\citeproc{ref-Kluyver2016Jupyter}{Kluyver et al., 2016}) with LaTeX
renderings and trajectory comparisons. It converts Antimony models to
SBML with the Antimony library
(\citeproc{ref-Heydarabadipour2026Antimony}{Heydarabadipour et al.,
2026}; \citeproc{ref-Smith2009Antimony}{Smith et al., 2009}), parses the
resulting SBML with libSBML
(\citeproc{ref-Bornstein2008libSBML}{Bornstein et al., 2008}), and
exposes direct SBML-file parsing through \texttt{ssys.parse\_sbml}.

\section{Statement of need}\label{statement-of-need}

S-systems and GMA systems are canonical ODE forms developed within
Biochemical Systems Theory (BST) (\citeproc{ref-Savageau1976}{Savageau,
1976}; \citeproc{ref-SavageauVoit1987}{Savageau \& Voit, 1987}).
Savageau \& Voit (1987) proved that systems built from sums, products,
and compositions of elementary functions can be recast into S-system
form; their theorem provides a minimum estimate of the class of
recastable systems. These canonical representations offer several
advantages: S-system steady-state equations become linear in log-space,
enabling algebraic analysis of identifiability
(\citeproc{ref-Villaverde2019Observability}{Villaverde, 2019}) and
sensitivity (\citeproc{ref-Savageau1971ParameterSensitivity}{Savageau,
1971}); the uniform power-law structure facilitates parameter estimation
via linear regression techniques (\citeproc{ref-Daniels2015b}{Daniels \&
Nemenman, 2015b}); and exact recast models can serve as ground truth for
validating structure-learning algorithms that infer governing equations
from data (\citeproc{ref-Daniels2015}{Daniels \& Nemenman, 2015a}).

Despite the theoretical utility of S-systems and GMA systems, no
general-purpose, open-source tool previously existed to perform exact
recasting of supported symbolic ODE models. \texttt{ssys} fills this gap
by providing:

\begin{itemize}
\tightlist
\item
  \textbf{Exact transformation.} The recast is algebraically equivalent
  to the original on the constraint manifold, verified by symbolic
  differentiation and numerical comparison.
\item
  \textbf{Broad applicability.} The tool handles rational functions,
  selected elementary functions including \texttt{exp}, \texttt{log},
  \texttt{sin}, and \texttt{cos}, and explicit time dependence through
  systematic lifting. \texttt{sin} and \texttt{cos} are represented with
  positive offset auxiliaries; logarithmic lifts require positive
  arguments, and all lifted variables used as power-law bases must
  remain positive. \texttt{ssys} does not automatically translate
  variables that can be zero or negative, so users must preprocess those
  models into the positive orthant before recasting.
\item
  \textbf{Standard formats.} Input/output in Antimony format ensures
  interoperability with SBML-based workflows and model repositories such
  as BioModels (\citeproc{ref-Malik2020}{Malik-Sheriff et al., 2020}).
\item
  \textbf{Validation infrastructure.} A three-test suite (symbolic
  Jacobian verification, pointwise numerical sampling, trajectory
  comparison) provides certificates of correctness.
\end{itemize}

The package enables researchers to convert published models into
canonical form for analysis, create ground-truth benchmarks for
structure-learning algorithms, and leverage the log-linear properties of
S-systems for algebraic steady-state analyses.

\section{State of the field}\label{state-of-the-field}

Antimony is a human-readable modeling language for systems biology, with
transparent interoperability with SBML. The S-system and GMA formalisms
are classical (\citeproc{ref-Savageau1976}{Savageau, 1976};
\citeproc{ref-SavageauVoit1987}{Savageau \& Voit, 1987};
\citeproc{ref-VoitKemp2025}{Voit \& Kemp, 2025}), but existing software
support has focused on analysis, construction, or approximation rather
than exact general-purpose recasting. Classical BST
computational-analysis workflows focus on systems already expressed in
S-system or GMA form (\citeproc{ref-Voit2000ComputationalAnalysis}{Voit,
2000}). PLMaddon (\citeproc{ref-Vera2007PLMaddon}{Vera et al., 2007})
can generate power-law approximations via Taylor expansion, but such
local approximations differ fundamentally from exact recasting with
auxiliary variables. More recently, BSTModelKit.jl
(\citeproc{ref-Vadhin2026BSTModelKit}{Vadhin \& Varner, 2026}) provides
Julia tools for constructing and analyzing BST models, though it
emphasizes model construction and analysis from declarative
specifications rather than recasting supported symbolic ODE models from
SBML/Antimony inputs. To the best of our knowledge, no existing
open-source tool performs exact algebraic recasting from supported
SBML/Antimony ODE models to S-system or GMA form with explicit
constraint-manifold handling and validation certificates.

\section{Software design}\label{software-design}

\texttt{ssys} is organized as a parser, symbolic transformation engine,
formatter, and validation layer. Antimony models are converted to SBML
with the Antimony library and then parsed with libSBML; SBML files can
be parsed directly with \texttt{ssys.parse\_sbml}. Both paths produce
SymPy (\citeproc{ref-Meurer2017SymPy}{Meurer et al., 2017}) symbolic
expressions, which serve as the internal representation for exact
algebraic rewriting.

The recaster applies local exact transformations: composite functions
(exp, log, sin, cos, etc.) and rational denominators are lifted to
auxiliary variables with chain-rule-derived ODEs, and sums of monomials
are factored into products via pool-auxiliary construction. The
simplified output mode allows for relaxed S-system form, whereas the
canonical mode rewrites equations to a strict two-term S-system form.
Inputs with unsupported SBML semantics, unsafe generated derivatives, or
violated domain assumptions are rejected, reported as unsupported, or
fail validation rather than being treated as validated recasts.
Correctness is checked by three independent validators: symbolic
Jacobian chain-rule verification, pointwise numerical sampling, and
trajectory comparison via libRoadRunner simulation
(\citeproc{ref-Somogyi2015libRoadRunner}{Somogyi et al., 2015};
\citeproc{ref-Welsh2023libRoadRunner}{Welsh et al., 2023}).

\section{Quality control}\label{quality-control}

The package includes 712 pytest-collected test cases covering parsing,
recasting, validation, and CLI functionality. Of these, 11 are
integration-marked test cases, including 8 slow manifest tests that
entail transformation and validation of 117 handwritten Antimony models
across two output modes. Most of these models were drawn from published
examples used to introduce, solve, or analyze S-system and related
power-law representations: the complete Savageau \& Voit example set
(\citeproc{ref-SavageauVoit1987}{Savageau \& Voit, 1987}), later
S-system examples and applications
(\citeproc{ref-Irvine1988SSystemSolution}{Irvine, 1988};
\citeproc{ref-RustVoit1990}{Rust \& Voit, 1990};
\citeproc{ref-Savageau1988IntroSSystems}{Savageau, 1988},
\citeproc{ref-Savageau1993Roots}{1993};
\citeproc{ref-Voit1988NewMethodologies}{Voit, 1988a},
\citeproc{ref-Voit1988Recasting}{1988b},
\citeproc{ref-Voit1990Endemic}{1990},
\citeproc{ref-Voit1992Symmetries}{1992},
\citeproc{ref-Voit1993ChaoticInput}{1993}), metabolic optimization and
model-inference examples using power-law canonical forms
(\citeproc{ref-Daniels2015b}{Daniels \& Nemenman, 2015b};
\citeproc{ref-MarinSanguino2007}{Marin-Sanguino et al., 2007};
\citeproc{ref-Pozo2011}{Pozo et al., 2011}), and nonlinear dynamics or
control examples that test polynomial and quasipolynomial
transformations (\citeproc{ref-Anghel2013Lyapunov}{Anghel et al., 2013};
\citeproc{ref-Anguelov2018Gompertz}{Anguelov et al., 2018};
\citeproc{ref-HernandezBermejo1998}{Hernández-Bermejo et al., 1998};
\citeproc{ref-PapachristodoulouPrajna2005}{Papachristodoulou \& Prajna,
2005}; \citeproc{ref-Zhang2022PLL}{Zhang et al., 2022}).

\section{BioModels database
benchmark}\label{biomodels-database-benchmark}

To assess applicability to real-world models, we applied \texttt{ssys}
to ODE models from the BioModels database
(\citeproc{ref-Malik2020}{Malik-Sheriff et al., 2020}). The benchmark
suite (included in the repository under \texttt{biomodels\_batch/})
operates in three phases: (1) fetch SBML models from BioModels,
filtering for ODE-based models; (2) identify transformation candidates
by excluding models with unsupported features or more than 100 dynamic
species, 200 reactions, or 500 parameters; and (3) batch transform the
candidates. After filtering 1,644 SBML models downloaded from BioModels,
978 candidates remained. Of these, the transformation algorithm
completed for 848 (86.7\%), and numerical validation confirmed
mathematical equivalence for 739 transformed models. Among the validated
models, 366 achieved meaningful structural canonization: 282 models in
general form were converted to GMA, 77 models in GMA form were converted
to S-system form, and 7 models in general form were converted to
S-system form. The remaining validated models were already in GMA or
S-system form, or retained general form after exact rewriting.
Unsuccessful cases were mostly due to unsupported model features,
parsing or loading errors, complexity limits, or timeouts. Benchmark
scripts and the tracked summary are in \texttt{biomodels\_batch/};
rerunning the benchmark writes detailed outputs under
\texttt{biomodels\_batch/results/}.

\section{Research impact statement}\label{research-impact-statement}

\texttt{ssys} makes recasting construction
(\citeproc{ref-SavageauVoit1987}{Savageau \& Voit, 1987}) available as a
practical tool for SBML and Antimony models. The curated handwritten
examples provide regression tests with known recasting targets, while
the BioModels benchmark demonstrates that exact recasting can be applied
at repository scale to hundreds of published ODE models.

\section{Availability}\label{availability}

\texttt{ssys} is available on GitHub at https://github.com/lanl/ssys.
Installation requires Python \textgreater= 3.10 (and \textless{} 3.13),
SymPy, NumPy (\citeproc{ref-Harris2020NumPy}{Harris et al., 2020}),
libRoadRunner, and the Antimony library. The repository includes
comprehensive documentation: theoretical background and recasting rules
(\texttt{RECASTING.md}), test model descriptions
(\texttt{TEST\_MODELS.md}), and usage instructions (\texttt{README.md}).

Installation: \texttt{pip\ install\ ssys}. For a single-model
command-line example, put the path to an Antimony file such as
\texttt{test\_models1/m01\_exp\_decay.ant} in a one-line manifest and
run
\texttt{ssys-recast\ -\/-manifest\ models.manifest\ -\/-outdir\ out\ -\/-validate}.

\section{AI usage disclosure}\label{ai-usage-disclosure}

Generative AI assistance was used during software review, debugging,
benchmark analysis, bibliography checking, and manuscript editing. The
author made the scientific, architectural, and editorial decisions,
reviewed AI-assisted changes, ran verification tests and paper builds,
and is responsible for the correctness and licensing of the submitted
materials.

\section{Acknowledgments}\label{acknowledgments}

This work was inspired by a phone call with Michael A. Savageau in early
2025. I will always be grateful to Mike for his kind and patient
mentorship of me as a graduate student at the University of Michigan,
his exemplary and inspiring scholarship, and his enduring support.

This work was supported by the National Institutes of Health (NIH)
National Institute of General Medical Sciences (NIGMS) under grant
R01GM111510 and by the U.S. Department of Energy through the Los Alamos
National Laboratory (LANL). LANL is operated by Triad National Security,
LLC, for the National Nuclear Security Administration of the U.S.
Department of Energy (Contract No.~89233218CNA000001).

\section*{References}\label{references}
\addcontentsline{toc}{section}{References}

\protect\phantomsection\label{refs}

\end{document}